\documentclass[twocolumn,showpacs,preprintnumbers,amsmath,amssymb]{revtex4-1}
\usepackage{graphicx}
\usepackage{dcolumn}
\usepackage{bm}
\usepackage{amsmath, amssymb, gensymb}
\usepackage{subfigure}
\begin{document}
\newcommand{\W}{7.5cm}
\newcommand{\ud}{\rm d}
\newcommand{\un}{~\mathrm}
\newcommand{\ie}{{\em i.e. }}
\newcommand{\eg}{{\em e.g. }}
\newcommand{\unm}{~\mu\mathrm{m}}
\newcommand{\indice}[1]{\textnormal{\scriptsize{#1}}}

\title{Large scale flow visualization and anemometry applied to lab on chip models of porous media} 
\author{Johan Paiola, {$^{1,2}$}}
\author{Harold Auradou,$^{1}$}\email{auradou@fast.u-psud.fr}
\author{Hugues Bodiguel ${2,3}$}

\affiliation{$^1$ Laboratoire FAST, Univ. Paris Sud, CNRS, Universit{\'e} Paris-Saclay, F-91405, Orsay, France.\\
$^2$ Univ. Bordeaux, CNRS, Solvay, LOF UMR5258, Pessac, France. \\
$^3$ Univ. Grenoble Alpes, CNRS, Lab. LRP UMR5520, F-38000 Grenoble, France.}

\begin{abstract}The following is a report on an experimental technique allowing to quantify and map the velocity field with a very high resolution and a simple equipment in large 2D devices. A simple Shlieren technique is proposed to reinforce the contrast in the images and allow you to detect seeded particles that are pixel-sized or even inferior to it. The velocimetry technique that we have reported on is based on auto-correlation functions of the pixel intensity, which we have shown are directly related to the magnitude of the local average velocity. The characteristic time involved in the decorrelation of the signal is proportional to the tracer size and inversely proportional to the average velocity. We have reported on a detailed discussion about the optimization of relevant involved parameters, the spatial resolution and the accuracy of the method. The technique is then applied to a model porous media made of a random channel network. We show that it is highly efficient to determine the magnitude of the flow in each of the channels of the network, opening the road to the fundamental study of the flows of complex fluids. The latter is illustrated with yield stress fluid, in which the flow becomes highly heterogeneous at small flow rates.
\end{abstract}

\maketitle

\section*{Introduction}

One of the challenges of microfluidic technology is the development of tools to characterize flow properties with a suitable resolution \cite{Stone2004}. For nearly twenty years, the PIV (Particle Image Velocimetry), the PTV (Particle Tracking Velocimetry) or the LIF (Laser Induced Fluorescence) methods have been successfully adapted to get information about the flow at the microscale \cite{Santiago1998,SINTON2004}. Recent developments have even enabled us to get 3D visualizations of the flows allowing an enriched description of the processes \cite{Cierpka2012,Kim2013}. However, these techniques are developed in order to have an accurate description of the flow field at the smallest possible scale - the single cell \cite{Zheng2013} or bacteria scale \cite{Drescher2011} for instance - but they only cover a limited range of scales. Complex microfluidic devices typically involve tens to hundreds of channels, and therefore ask for a velocimetry, or at least, anemometry technique having a super-high spatial resolution.

In the past decades, micromodels obtained using transparent microfluidics technology have been widely used \cite{CRANDALL2008,PERRIN2006,COTTIN2010,Clarke2015,conn2014,ma2012,LEE2015,ROMANO2011,SONG2015,BEAUMONT2014,Karadimitriou2014,WU2012,GUNDA2011,ferno2015} to model flows in porous media. Because, this technique allows to study at the relevant length scales and with a precise control of the geometry flow in porous media and it has permitted significant progress on the understanding and on the description of the flows in those media. However, characterizing flows in micromodels represent a technical challenge, as one need to perform measurements over length scales ranging from the pore scale up to a scale for which the macroscopic properties of the porous media are reached. For porous media modelled by a disordered channel network, it is assumed that a typical number of a hundred parallel channels is a minimum value to be statiscally representative. Under this condition, a single pore only represents from 0.1 to 0.5\% of the full scale. As few pixels per pore are necessary to characterize the flow, the resolution rapidly reaches the limit of standard cameras: a few thousands pixels per dimension. Therefore, flow characterization requires a very good resolution over a wide ranges of scales, which is in practice difficult to achieve. This issue is likely to be the reason why most of the past studies either were restricted to qualitative observations\cite{ma2012,ferno2015}, either were limited to the measurement of the saturation \cite{ROMANO2011,conn2014}, or were considering flow at the scale of a few pore\cite{Clarke2015}. Only a few studies report quantitative velocity measurements \cite{COTTIN2010,BEAUMONT2014}. Yet, they toke advantage of the good contrast existing in biphasic flows. Yet, this technique gives solely access to menisci velocity, and does not give information about the flow inside the fluids. 

The purpose of this work is go beyond these limitations, and to propose an anemometry technique having an extra-high resolution. Although it has been developed to tackle the technical issue of mapping the velocity in a micromodel of porous media, the technique is versatile and could be directly applied to complex microfluidics networks. 

When the acquisition device does not capture the full image, one solution is to combine a mosaic of images to reconstruct the complete velocity field. Yet, this procedure is only adapted to steady flows and requires the use of adapted equipments and of a reconstruction software to combine the images. In this work, by pushing the resolution limit two its minimal value: the pixel size, we were able to avoid reconstruction procedures. To do so, we use the time variation of the light intensity on each pixel rather than space correlation of successive images. The decay of the autocorrelation function of the light intensity is - in the high P{\'e}clet limit - related to the local velocity. As compared to PTV or to PIV, only the magnitude of the velocity can be measured. In a network, velocity direction is known a priori from channel orientation, so that this disadvantage is not highly relevant in these applications. The main advantage of time autocorrelation is that it does not require particle detection as in PTV, neither uniform and good quality images as in standard PIV. We show in this work that it could be applied on images of poor quality, where particle displacement is hardly seen by eyes. 

The second issue is to use tracers that do not perturb the flow: particle size should thus not exceed a tenth of the channel width. On the other hand, tracers have to be in the order of the pixel size. This last requirement makes particle detection difficult. One solution is to use fluorescent particles, but their use at the centimeter scale or above is not straighforward, and is not accessible by standard fluorescence microscopy equipments. By taking advantage of the Schlieren effect\cite{Sutherland1999,Dalziel2000}, our technique avoids the use of fluorescent tracers and only required a standard bright field illumination. The Schlieren effect is obtained by simply placing a grid between the device and the light source. The grid enhances the contrast of the images allowing particle detection even if the particle size is below the pixel size. Despite the contrast enhancement, the images are of poor quality but sufficient for a time autocorrelation analysis of the light intensities.

In this paper, we present in detail the principles and the validation of the autocorrelation particle imaging anemometry. The technique is applied to map the velocity field in a complex and wide microfluidic network of channels. We show that the method is simple, fast and importantly leads to an high-resolved determination of the mean velocity in all the channels simultaneously. These results and validation are obtained for Newtonian fluids. In order to illustrate the potential use of the technique, we present some results obtained with a the yield stress fluid.

\section*{Description of methods and materials}
\label{sec:mat}
\begin{figure*}[t]
 \centering
 \includegraphics[width=.8\linewidth]{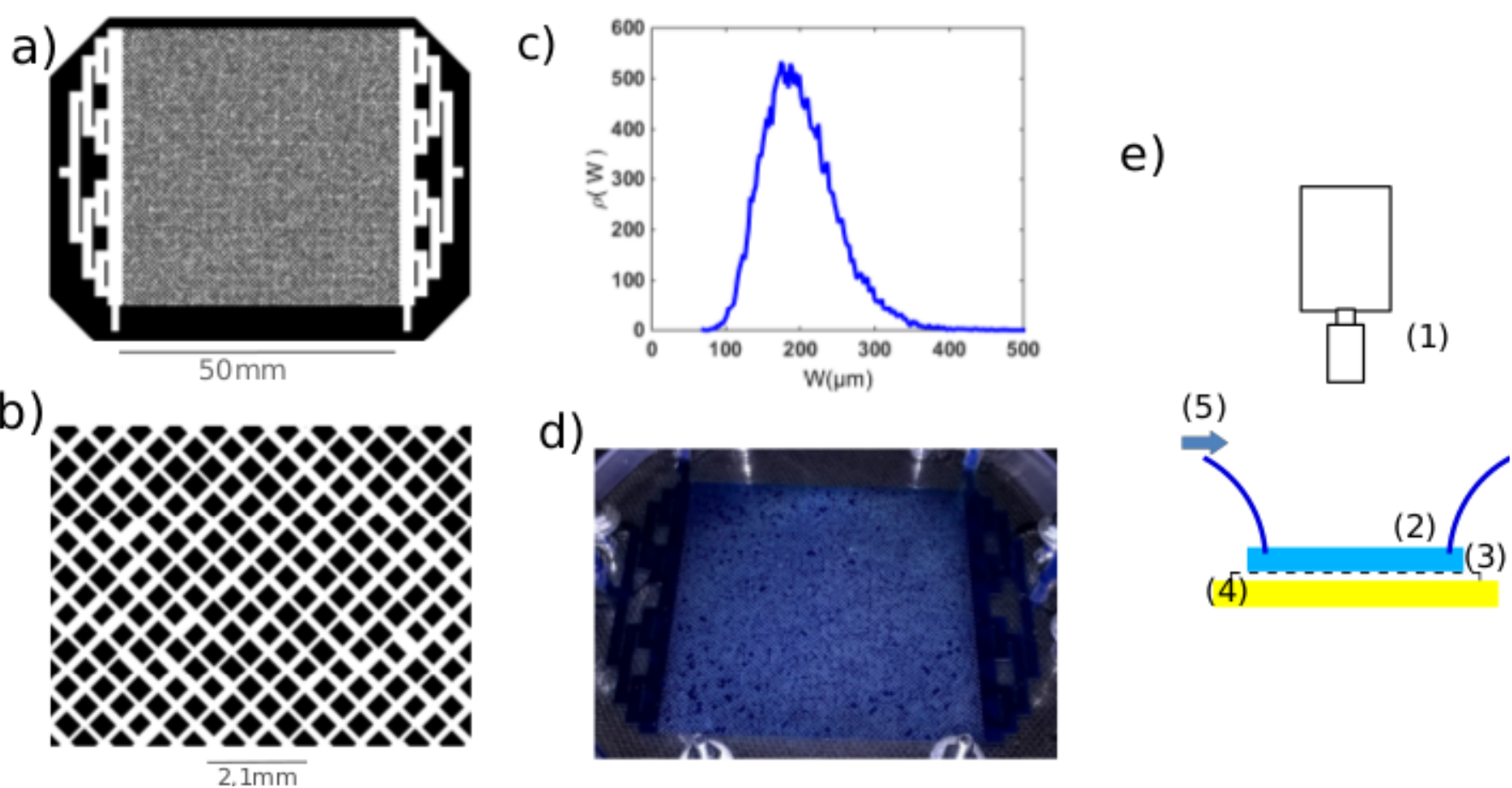}
\caption{(a) mask used for the lithography. The microfluidic chip consists of a 2D network made of $110$ by $110$ channels. The average channel width ($W$), length ($l$) and depth ($h$) are respectively: $W=200$, $l=700$ and $h=600 \mu m$. (b) Close-up view of the network. (c) Size distribution of the width $W$ of the channels. (d) Photograph of the assembled microfluidic device filled with a dyed fluid. (e) Schematic view of the experimental set-up: 1- camera with a $12.5 mm$ lens, 2- microfluidic chip, 3- grid with a square mesh, 4- light panel, 5- connections. }
 \label{fig:fig1}
\end{figure*}

\label{sec:method}
\subsection*{System overview}

Two microfluidic devices were developed using classical soft lithography technique. 

The first one is a channel of constant depth ($h=340 \mu m $) but of gradually increasing width. The widths are respectivelty $W=150$, $200$, $300$, $400$ and $500 \mu m$ and the total length of the channel is 10 $mm$. This device is used to calibrate and validate the autocorrelation technique at the pore scale. 

The second one, displayed in Fig. \ref{fig:fig1}, is a square lattice of $110$ by $110$ channels of rectangular cross sections. The height of the channels is uniform equal to 200 $\mu$m, while their widths are set randomly according to a log-normal distribution of mean value 200~$\mu$m and standard deviation 50~$\mu$m (see fig. \ref{fig:fig1}b). The lattice is orientated at 45$^\circ$, and $N=110$ channels are connected to the inlet. A tree-like geometry made of channels of width 2000 $\mu$m is used at the inlet and at the outlet to impose a parallel flow (see Fig. \ref{fig:fig1}a).

The inlet of the device is connected to a syringe pump and the outlet to a waste reservoir. We used either pure glycerol or yield stress fluid obtained by mixing $0.7\ g$ of carbopol EDT2050 in water (See reference \nocite{Geraud2013}\citenum{Geraud2013} for a complete description of the preparation procedure). The flow curve of the carbopol solution has been determined using standard rheometry and is well described by a Hershel-Buckley law: $\sigma =15 + 5.1\dot{\gamma}^{0.5}$. The fluids are seeded with transparent beads of PMMA. Two diameters are used: $6$ and $20 \mu m$. The concentration is $0.5\%$ for the bead $6 \mu m$ and $0.3\%$ for the bead $20 \mu m$.

A JAI BM 500 GE video camera is positioned above the setup. Its CMOS sensor consists of $2456$ by $2058$ pixels and its maximum rate of capture is $7$ frames per second in full resolution. The spatial resolution of the camera has been varied to test the limitation of the technique, but for the anemometry mapping of the channel network, it is about 25 $\mu $m per pixels. This resolution doesnot permit to visualize and track single particles, but, as detailed in the next section, contrast enhancement by a Schlieren technique allow to detect intensity fluctuations due to tracers displacement.

\subsection*{Contrast enhancement}

\begin{figure}[h]
 \centering
 \includegraphics[width=\linewidth]{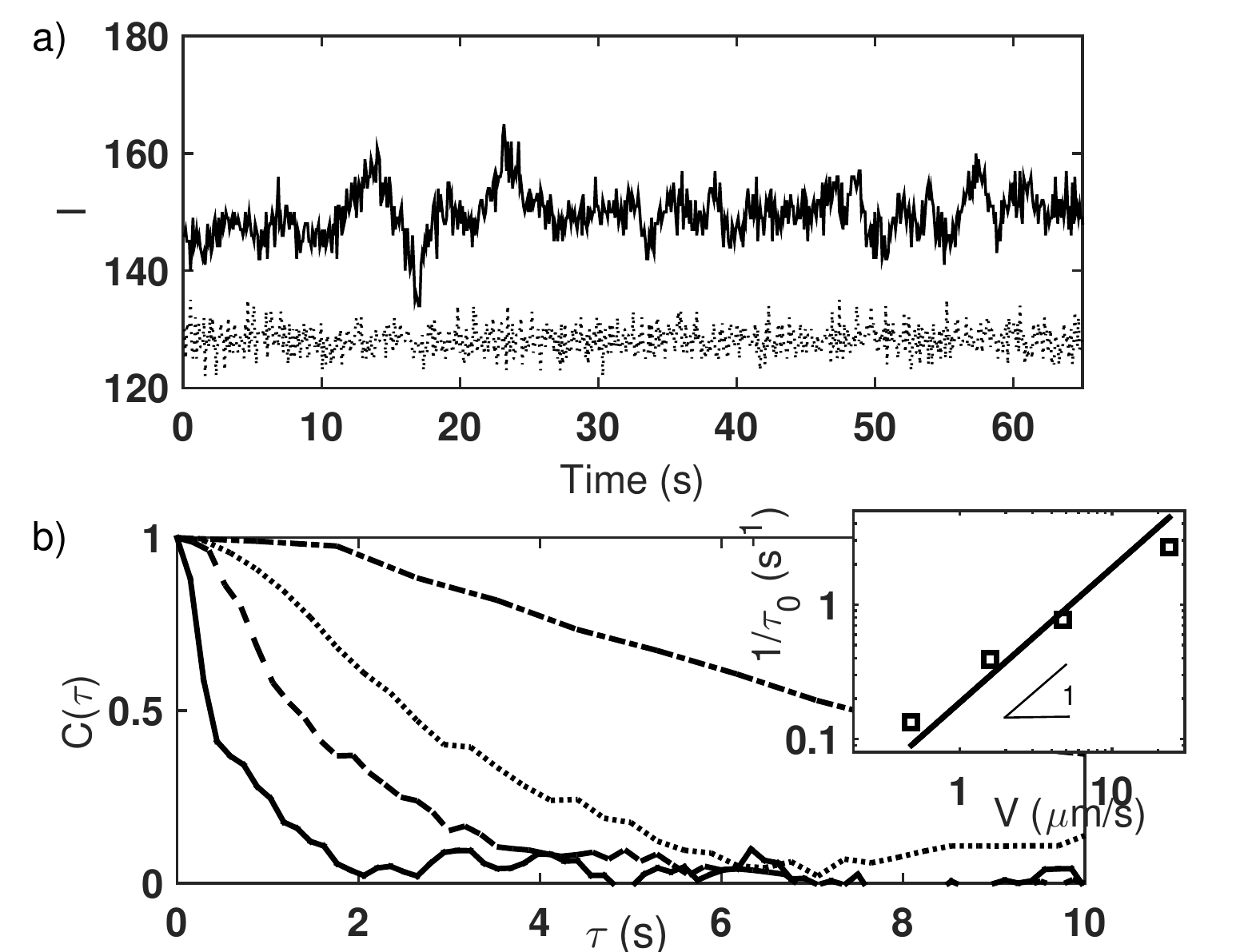}
 \caption{\textit{a)} Variation of the light intensity $I(x,y,t)$ with time. Solid and dotted lines are measurements done with $V= 4.8 \mu m.s^{-1}$ respectively with and without the grid. 
\textit{b)} Auto-correlation function $C(\Delta t)$ in relation with the time lag $\Delta t$ for $V=23.8$ (solid line), $4.8$ (dashed line), $1.6$ (dotted line) and $0.48 \mu m.s^{-1}$ (dash-dotted). For these experiments, the beads diameter is $20 \mu m$ and we used a grid of $40 \mu m$ mesh size. Spatial resolution is $22 \mu m /pixel$. In insert the characteristic time $\tau_0$ is plotted as a function of the mean velocity (see text).}
 \label{fig:fig3}
\end{figure}

To enhance the contrast we adapted the synthetic Schlieren technique \cite{Sutherland1999} developed in the 90's to measure density fluctuations in two-dimensional stratified flows. In that case, density fluctuations are revealed through small optical refraction index variations enhanced by placing a grid between the light source and the observation field. Our technique uses the refraction index contrast between particles suspended in the fluid and the fluid. Schlieren techniques also require an optimized light. Here, this is achieved by placing a grid between a light panel and the device. The grid produces a uniform field of dotted sources of light and the light passing through the beads is refracted, thus reducing the light intensity at that position. The grid is a sieve of mesh sized 20 or 40 $\mu$m. By doing so, we immediately observe light intensity fluctuations due to particle displacement, as illustrated in Fig. \ref{fig:fig3}a. These fluctuations are hardly distinguishable from noise in the absence of the grid (dotted line in Fig. \ref{fig:fig3}a) but are enhanced when a grid is placed between the light and the device (solid line in Fig. \ref{fig:fig3}a). 

\begin{figure}[h!]
 \centering
\includegraphics[width=\linewidth]{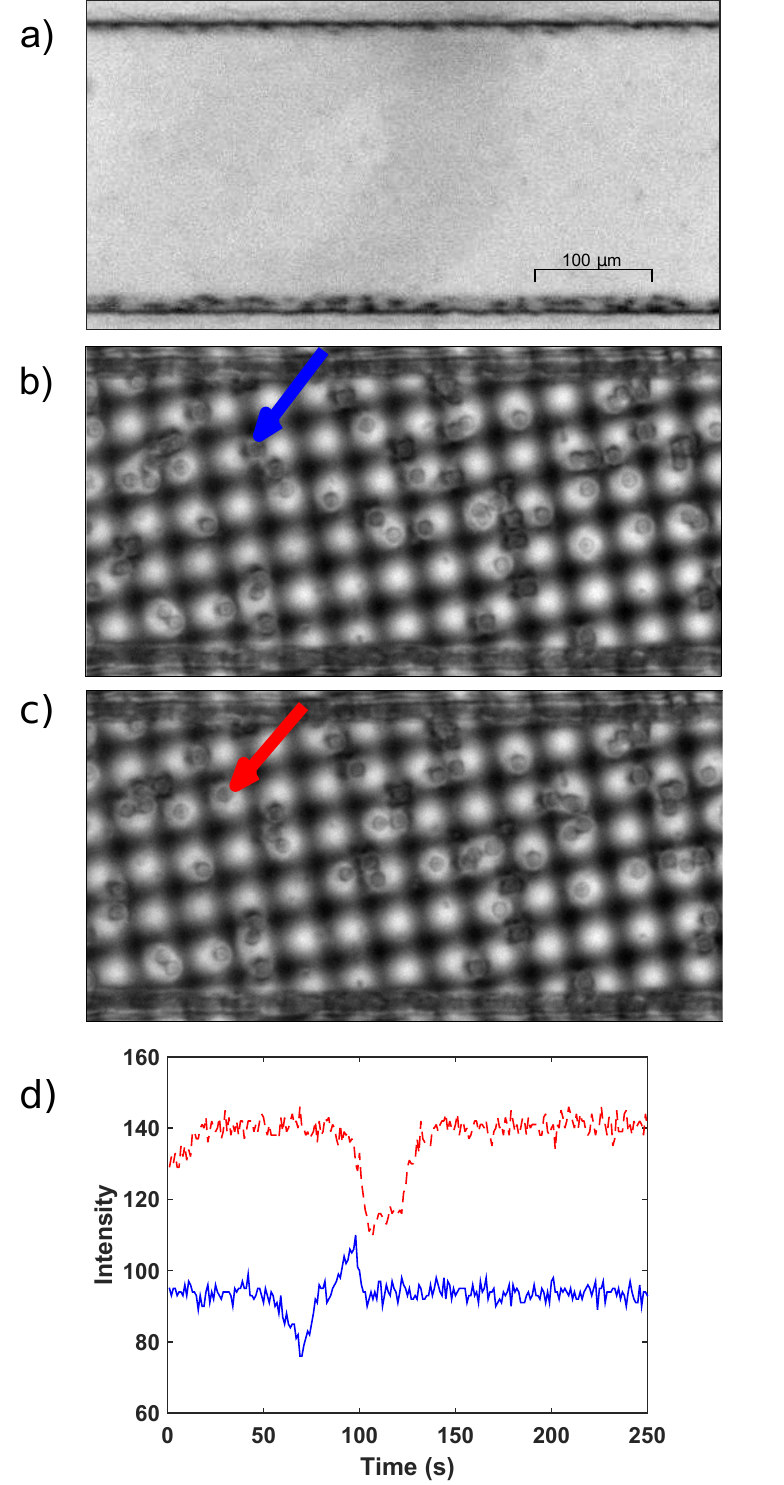}
\caption{(a) close-up view of a single channel. The width of the channel is $250 \mu m$. The channel is filled with glycerol containing PMMA beads of diameter $20 \mu m$. (b,c) Same view but with a grid added between the light pad and the microfluidic device. The mesh of the grid is $40 \mu m$. The grey circles are beads.(d) Light intensity measured on two particular pixels in relation with time. The dotted red and solid blue lines are the signals recorded for respectively the beads shown by the red (b) and blue (c) arrows.
The spatial resolution is $1.15 \mu m/pixel$ and the mean flow velocity $V=16 \mu m.s^{-1}$.}
 \label{fig:fig2}
 \end{figure}

Before demonstrating the correlation between the light intensity variations and the displacement of the beads, it worth analyzing in more details the contrast enhancement due to the grid. For that purpose, we approached the camera to get magnified views of the particles. Fig.~\ref{fig:fig2} displays examples of pictures obtained with and without the grid. The contrast difference is striking, as the beads become highly visible when the grid is present. This contrast enhancement requires that the grid is placed close to the observation plane, and that the mesh wavelength is in the order of the that of the beads. The physical origin of this effect is related to light refraction by the beads. Without the grid, the diffuse nature of light prevents a good contrast. The grid creates an array of point light sources and deviations of light beams directly modifies the light intensity in the field of view.

In the presence of a flow, particle displacement leads to intensity modulation on a given pixel. Examples are displayed in Figure \ref{fig:fig2}d. We note that the intensity changes depend on the location. When it is calculated on a bright pixel taken inside a hole of the grid, the intensity is lower when a particle is present. However, when the pixel is chosen in a darker area, the modulation is more complex. In the example shown in Fig. \ref{fig:fig2}d, the intensity is first lower and then higher than the baseline. This comes from complex light refraction by the beads in the vicinity of the meshes. 

When the camera is moved backwards, the detail of the diffraction pattern due to the grid is no more visible but the light intensity contrast between the bead and the fluid is still sufficient to be seen with this lower spatial resolution. The complex intensity modulations which depends on the pixel location using a magnified view are no more visible in the sense that when the pixel size is about that of grid wavelength, all the pixel are equivalents. The final signal - as the one displayed in Fig.\ref{fig:fig3}a - is composed of a succession of individual signals corresponding to the passage of beads on and between meshes.

\subsection*{Image treatment}

To demonstrate that the fluctuations observed with a grid are due to the passage of the beads on that pixel, we compute the normalized time auto-correlation function of the centered intensity of single pixel. It is defined by
\begin{equation}
C(x,y,\Delta t) =\frac{\sum_t \left[I(x,y,t)- \left\langle I(x,y)\right\rangle _{t}\right]\left[I(x,y,t+\Delta t)-\left\langle I(x,y)\right\rangle_t\right]}
{\sum_t \left[ I(x,y,t) - \left\langle I(x,y)\right\rangle_t \right]^2},
\label{eq}
\end{equation}
where $\Delta t$ is the time lag and $I$ the pixel intensity. By construction, these functions decay and eventually vanish at long times. The correlation functions are calculated thanks to Fourrier transforms of the intensity, since the computation time is significantly decreases. Examples are displayed in Fig. \ref{fig:fig3}b for different flow rates. Clearly, the autocorrelation function decreases faster for high flow rates.	

The anemometry technique that is proposed in this article is based on this observation, and consists in measuring the characteristic time of the autocorrelation function decay. The latter is estimated as the time $\tau_0$ at which $C(\tau)=0.5$, and we use linear interpolation of the autocorrelation function to get a better time resolution. The inset in Fig. \ref{fig:fig3} shows the value of $1/\tau_0$ obtained from the measurements of the correlation functions for different mean velocity $V$ in the channel, defined by $V=Q/S$, where $Q$ is the flow rate and $S$ the channel cross-section. A linear relation between the two quantities is observed suggesting it is possible to obtain a local measurement of the fluid velocity from the measurement of the autocorrelation function of the light intensity. 



\section*{Results}
\label{sec:results}

\subsection*{Flow in a single channel}

\begin{figure}[h!]
 \centering
 \includegraphics[width=\linewidth]{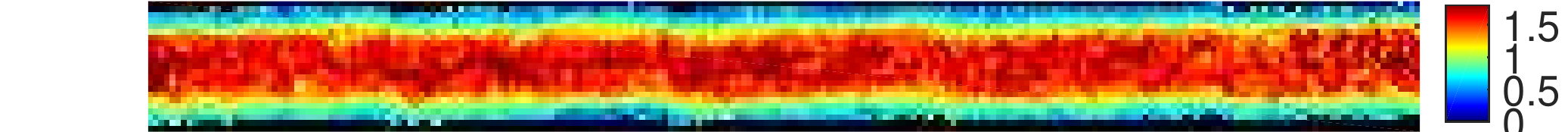}
\includegraphics[width=0.8\linewidth]{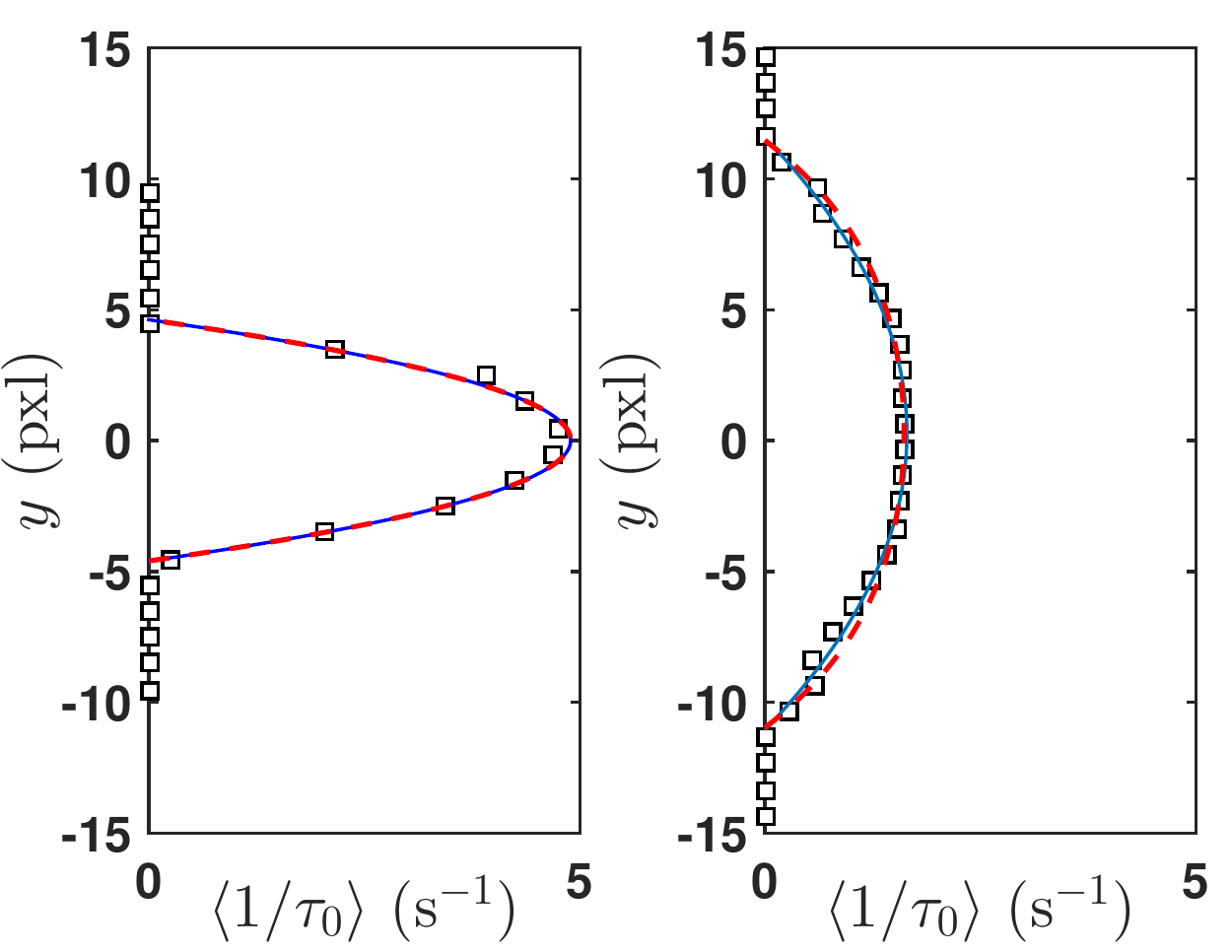}
 \caption{\textit{Top:} Color map of $1/\tau_0(x,y)$ measured on each pixel of a channel of width $500 \mu m$ and length $10 mm$. $V=1.3 \mu m.s^{-1}$. \textit{Bottom:} profiles of $\left\langle 1/\tau_0(x,y)\right\rangle_x$ where $x$ is the coordinate along the flow direction and $y$ the coordinate normal to the flow. The left and right figures show the result for channels of width $150$ and $500 \mu m$ respectively. The spatial resolution is here $15 \mu$ m/pixel. The diameter of the beads is $6\mu$m and the grid mesh size is $20 \mu$m. The solid lines are the best parabolic fit to the data, and the dashed lines corresponds to the expected velocity profile calculated for channels of rectangular cross-section, averaged over the depth of the channel. }
 \label{fig:fig4}
\end{figure}

We first applied the technique to experiments performed in straight channels. Fig.~\ref{fig:fig4} shows one of the maps of the inverse of $\tau_0$ that we obtained. At the center of the channel $\tau_0$ is small indicating the beads travel faster in the central region of the flow. On the sides, the time is large (and $1/\tau_0$ is small); in these regions the beads travel more slowly. From these maps, we estimate the velocity profiles by averaging $1/\tau_0$ along the flow direction. The profiles are shown on Fig. \ref{fig:fig4}b. They are very well adjusted over the full width of the channels by a parabolic profile as expected for Poiseuille flows in slits. It should be noted that the slit approximation is not strictly valid since the channel cross-section is rectangular, with an aspect ratio ranging between $W/h=0.44$ and $1.5$\footnote{In a channel of rectangular cross-section, the flow profile is given by\cite{white} $v(y,z)=\sum_{n=1,3,5...} (1- \cosh(n\pi y/h)/\cosh(n \pi w/2h)\cos(n\pi z/h)/n^3.$}. If we assume that the measured velocity is averaged over the entire thickness of the channel, the difference between the parabolic profile in a slit and the averaged profile in a rectangular channel is rather small. The two are are displayed in Fig. \ref{fig:fig4}b, and the difference remains smaller than the measurement uncertainty. Near the surfaces, fluctuations of the inverse of the time are larger, emphasizing the difficulty to achieve good measurements close to a surface. The fact that we used beads that are smaller than the pixel size allow to achieve spatial resolution of 1 pixel: as shown in Fig.~\ref{fig:fig4}b, a remarkable velocity profile is measured in a channel which is only 10 pixels wide.\\ 

\begin{figure}[h!]
	\centering
\includegraphics[width=.9\linewidth]{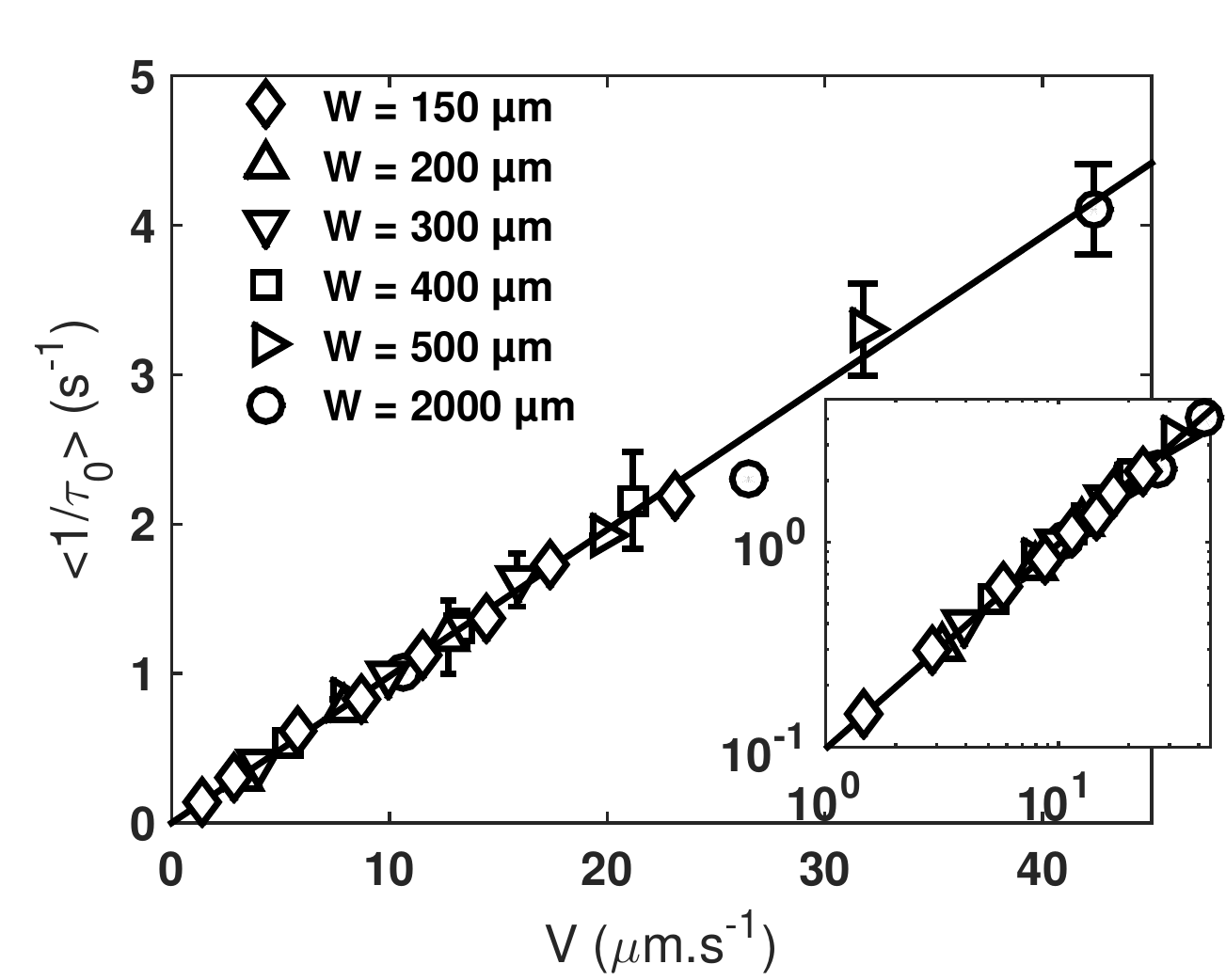}\\
\caption{Each symbol represents the average value of $1/\tau_0$ estimated from each experiment (different flow velocity $V$ and channel widths $w$) in relation to the average flow velocity $V$. The solid line is the linear regression of the measurements. Its slope is $1/a$ with $a=10.18 \pm 0.25 \mu m$. The spatial resolution here is $15 \mu m /pixel$. The diameter of the beads is $6\mu m$ and the grid mesh size is $20 \mu m$. The diamonds are the measurements performed in the injection channels of the network displayed on Fig.\ref{fig:fig6}. The conditions are the same (bead diameter $6\mu m$ / mesh size $20 \mu m$) except for the spatial resolution $25 \mu m /pixel$ instead of $15$. }
\label{fig:fig5}
\end{figure}

The experiment was repeated for different flow rates and the average value of $1/\tau_0$ was calculated on all the sub-parts of constant width of the channel. Fig. \ref{fig:fig5} shows the average value of $1/\tau_0$ in relation to the corresponding average flow velocity. The data fall on a single line passing through the origin. The slope $1/a$ is used to determine the relation between the time $\tau_0$ measured and the current velocity. We find a value of $a$ which is about 10 $\mu$m. This length is in the order of both the bead size and the pixel size, this will be discussed later. This result proves the ability of the auto-correlation technique to be used to measure local flow velocity, after calibration. This study also allows us to estimate the experimental uncertainty in the measurements of the magnitude of the velocity (approx. $5\%$).

We find that the calibration factor $1/a$ does not depend on the aspect ratio of the channel, as evidenced in Fig. \ref{fig:fig5}. Note that we have significantly varied the channel aspect ratio $W/h$ from 0.4 to 3.3. This indicates that the local velocity measured in one pixel is averaged along the channel depth. If the measurement was sensitive to tracer located in the center (for example), the calibration factor would have been different. Therefore, we conclude that the auto-correlation technique could serve as a local anemometry technique, which is of great interest in complex microfluidic networks since only a few pixels per channel is needed to achieve a precise measurement. It might also be extended for velocimetry application, but one would need to achieve a better optical vertical resolution using for instance confocal microscopy.

\subsection*{Flow in a complex channel network}

We next applied the method to the network of channels of random width depicted in Fig.~\ref{fig:fig1}. In this case, the spatial resolution is $25 \mu m$ per pixel and the goal is to measure the velocity over the full size of the chip. The spatial resolution is four times the beads diameter, which means they are not visible, preventing any tracking. Moreover, each channel contains in average only ten pixels, making it difficult to determine a flow profile precise enough for a good estimation of the average flow velocity using the classical techniques. 

\begin{figure}[h!]
\includegraphics[width=\linewidth]{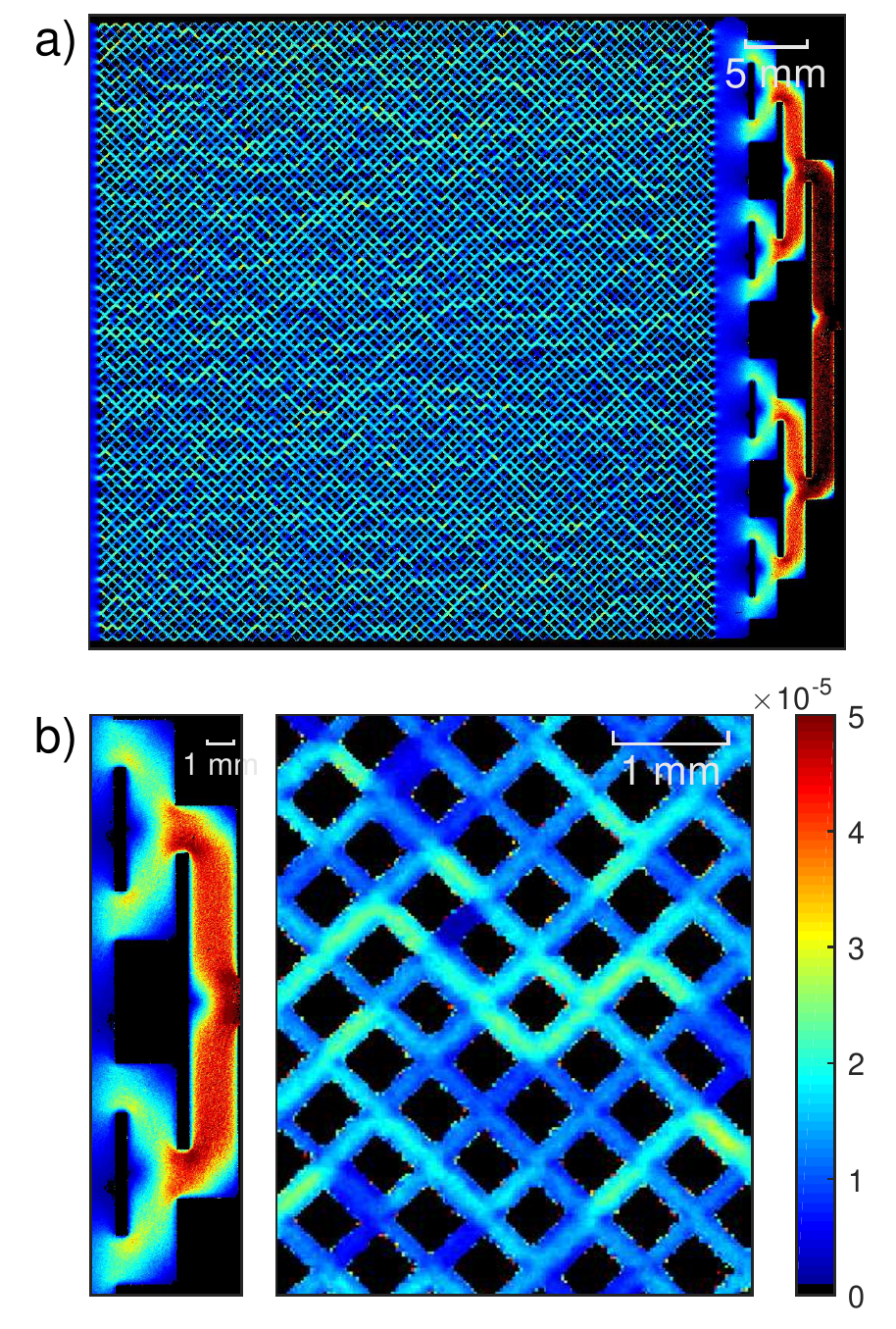}
\caption{\textit{a)} Color scale map of the flow velocity measured on each pixel. Pure glycerol containing beads of diameter $6 \mu m$ is injected at an average flow velocity $V=6.3 \mu m.s^{-1}$ in the micromodel. For this experiment, we used a grid of $20 \mu m.s^{-1}$. The spatial resolution is 25 $\mu$m/pixel. 
\textit{b)} Close up of the injection section and of a part of the network.}
\label{fig:fig6}
\end{figure}

\begin{figure}[h!]
\includegraphics[width=\linewidth]{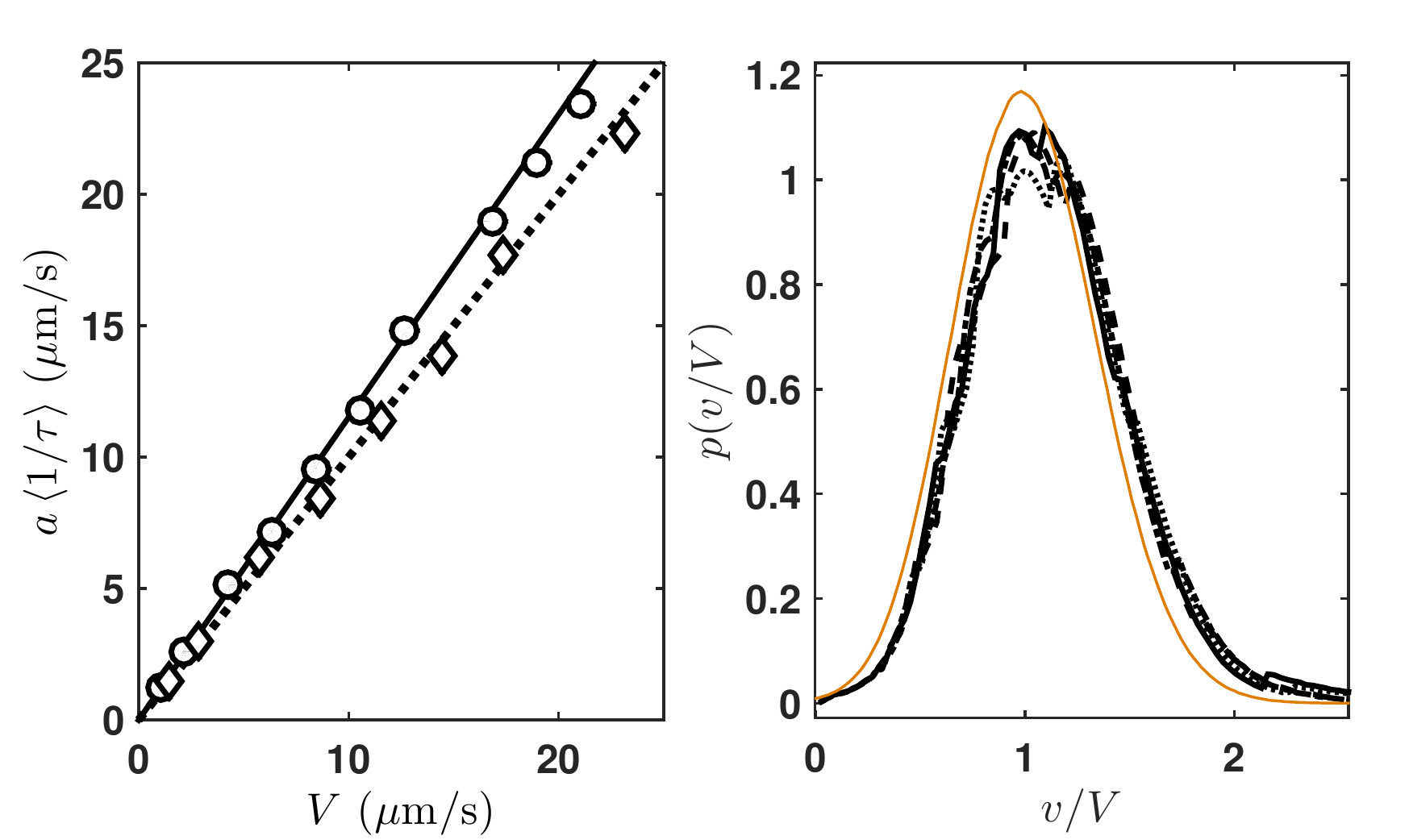}
	\caption{ \textit{Left:} Mean velocity in the porous medium (circles) determined for various flow rates $Q=NSV$. The calibration factor $a$ comes from the single channel experiment. The data are averaged over all the pore space of the micromodel. The solid line, of slope 1.15, corresponds to the best linear fit to the data. The slope of the dotted line is 1. Diamonds correspond the averaged velocity in the injection channels. \textit{Right:} Probability density functions of the normalized velocity in the porous medium. The thin solid line corresponds to the pore network simulations. It is well adjusted by a gaussian function of mean value 1 and of standard deviation 0.34. The other lines (dotted, dashed, solid and dash-dotted) are obtained from experiments carried at various flow rates ($1.1 \times 10^{-10}$ m$^3$/s, $1.4 \times 10^{-10}$ m$^3$/s, $2.2 \times 10^{-10}$ m$^3$/s and $2.8 \times 10^{-10}$ m$^3$/s, respectively). }
	\label{fig:fig6b}
\end{figure}

Despite these obstacles, we were able to obtain a map of the magnitude of the velocities with a good resolution (see Fig.\ref{fig:fig6}, using the calibration displayed in Fig. \ref{fig:fig5}. To quantitatively test our estimation of the flow velocities, we performed experiments at different flow rates, and computed the mean value of the velocity in the entire network. The results are shown in Fig. \ref{fig:fig6b}a. As for the single channel, the average flow velocity obtained by our technique varies linearly with the average imposed flow velocity $V=Q/NS$, where $N$ is the number of channels and $S$ the average cross-section. We note that the mean measured velocity is about 15\% higher than $V$. This over-estimation is not measured when the analysis is performed on the straight channels connecting the pump to the network (See Fig.\ref{fig:fig6b}a) and we obtained a satisfying agreement between these measurements and the calibration curve obtained from the measurement in straight channels (See diamonds in Fig.\ref{fig:fig5}). Several arguments might be proposed to explain the over-estimation of the fluid velocity. First, the velocity $V=Q/NS$ does not account for the contribution of the nodes of the network, while they represent about 20\% of the total volume. Since one node connects two channels, the velocity in the nodes is higher and the mean velocity measured is then higher than $V$. Second, we cannot exclude that the poor resolution prevent correct measurements of the low flow velocity near the walls, which would also lead to an overestimation of the flow velocity.

To make the potential of our method even clearer, we made a close up of the inlet sections and of a small part of the network, shown in Fig.~\ref{fig:fig6}b. 
The inlet consists of straight channels dividing into two equal channels in order to distribute the fluid equally over the full length of the chip. This part is thus made of branches with right angle corners: the method clearly highlights the low flow region appearing in this area. Likewise, in the area where the flow divides into two, the "jets" in the vertical walls can easily be distinguished. On the scale of a few pores, we see that the method makes the velocity contrast between the pores visible.\\ 

We also compared the velocity probability density functions (PDF). Due to the Newtonian nature of glycerol, once normalized by the average flow velocity $V=Q/NS$, the distribution should remain unaffected by any change in the average flow velocity. The results are shown in Fig.\ref{fig:fig6b}b. The collapse of the PDF obtained at various flow rates is excellent, which demonstrates the good quality of our measurement. The PDF are well described by gaussian functions of standard deviations around 0.35. The experimental results are compared to numerical simulation of the flow using a pore-network model approach. This method consists in assuming a developped laminar Poiseuille flow profile in each of the channels, and to solve the linear system of equations obtained by writing mass conservation in each node of the network (see appendix for details). This allows to calculate the PDF of the velocity for the network used experimentally. The PDF is displayed in Fig. \ref{fig:fig6b}b, together with the experimental data. As already discussed, the experimental mean value is slightly higher than the theoretical one, but one could see that the agreement on the standard deviation of the PDF is excellent.

We conclude that the auto-correlation technique combined with contrast reinforcement is well adapted to characterize flows in complex geometries.


\subsection*{Flow of a non-Newtonian liquid in a complex channel network}

\begin{figure}[h!]
	\centering
 \includegraphics[width=\linewidth]{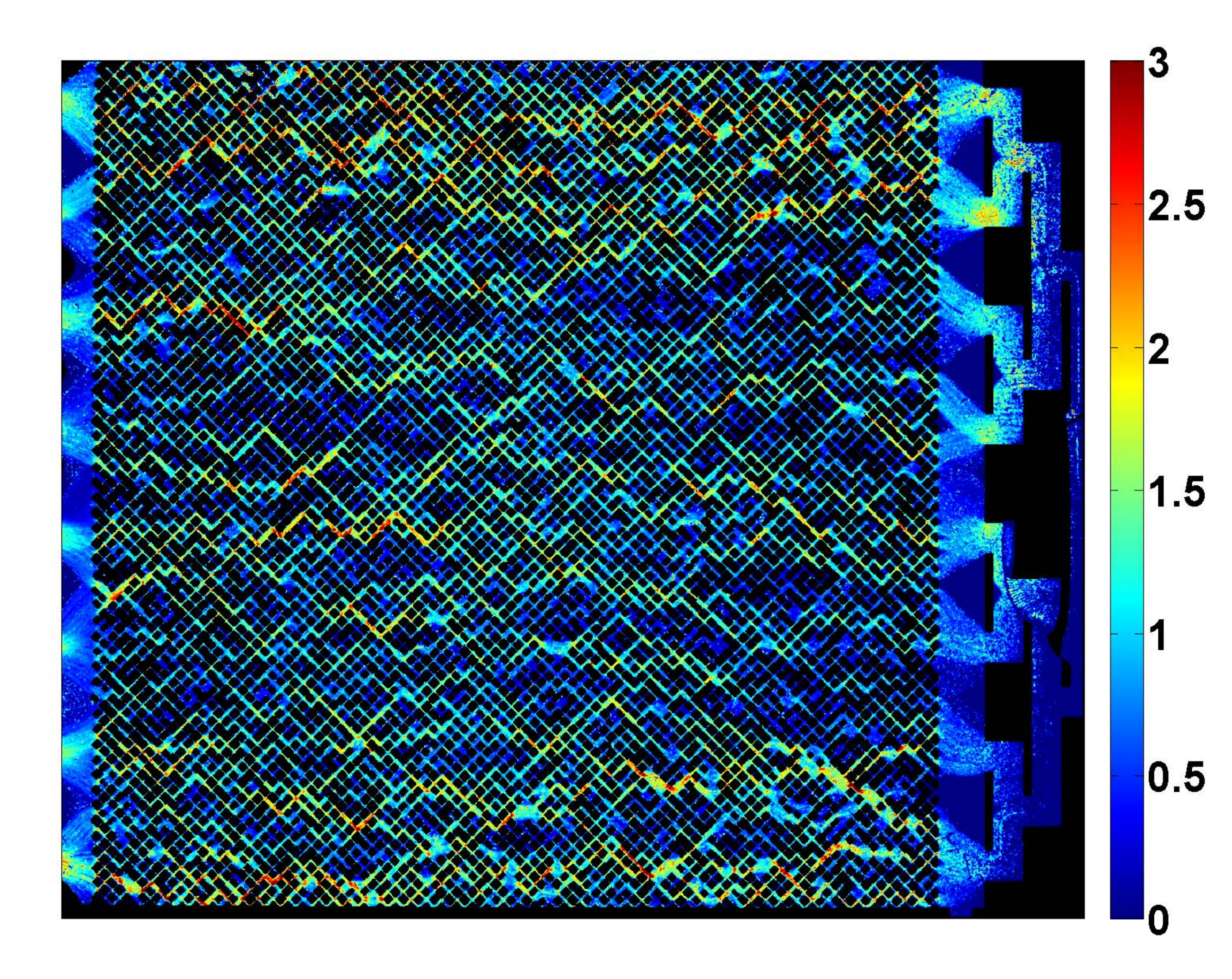}
\includegraphics[width=\linewidth]{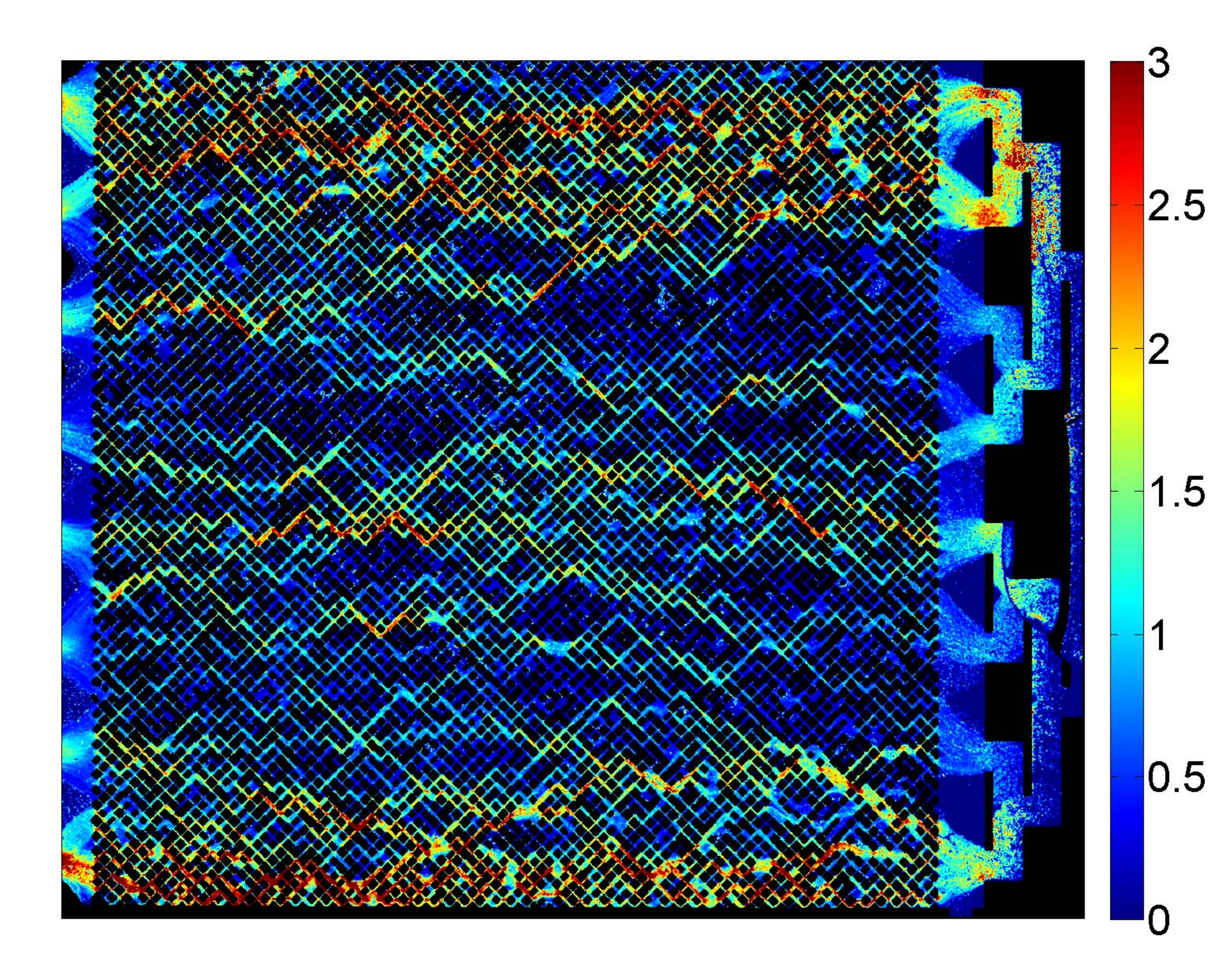}
\includegraphics[width=\linewidth]{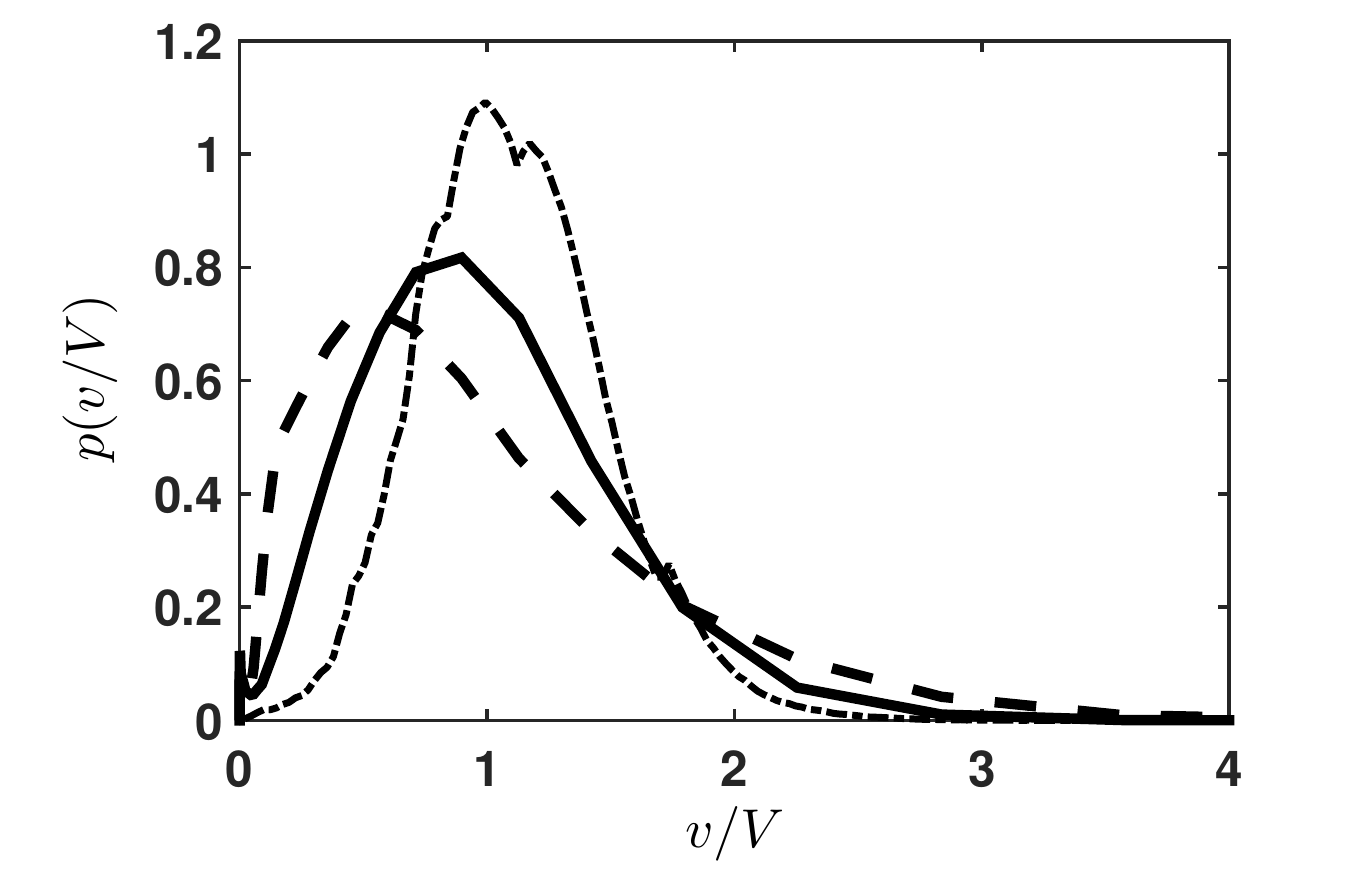}
	\caption{\textit{Top and middle: }Velocity maps obtained with a aqueous carbopol solution. The two maps correspond to two different degrees of pressure applied $\Delta P=7000 $~Pa (averaged velocity: $V=3.2 \mu m.s^{-1}$) and $5000\ Pa$ ($V=1.4 \mu m.s^{-1}$). \textit{Bottom:} distributions of the normalized velocity $v/V$ in the two experiments shown above (solid and dashed line) and of a Newtonian fluid (dotted line). }
\label{fig:fig8}
\end{figure}

In order to illustrate the potential interest of our method, we used it to determine the flow structure of a complex fluid in the micromodel. Contrary to Netwonian fluids flows which could be easily predicted or modeled, characterizing the flow of non-Newtonian fluids in complex geometries is far more challenging because of the strong coupling between the geometry and the fluid properties. We choose to use a yield stress fluid, a carbopol gel. The main characteristic of this fluid is the existence of a yield stress $\sigma_y$ (approx. $15\ Pa$). Consequently the flow in a single channel only occurs when the pressure gradient at the pore scale $\Delta p /l$ overcomes the strain on its surface $2\sigma_y/W$. In a network of heterogeneous channels, one thus expects that for low flow rates, the flow will is localized along a few paths, the other channels being below the yield stress. The existence and characteristics of preferential paths have been predicted numerically\cite{talon2012,Chevalier2015} but yet not tested experimentally. One of the practical consequence of these preferential paths is to affect the global pressure drop/flow rate relation. The current scientific challenge is to determine the equivalent Darcy law for yield stress fluids. While most of present studies focus on the macroscopic average flow velocity \cite{Chevalier2013}, we see on Fig.\ref{fig:fig8} that our method makes it possible to study the details of the local flow structure. 

The two velocity maps displayed on Fig.~\ref{fig:fig8} are obtained with pressure $\Delta P$ close to the theoretical macroscopic yield stress pressure. As compared to the flow of a Newtonian fluid displayed in Fig. \ref{fig:fig6}, the flow of the carbopol solution is much more heterogeneous. They both reveal the localization of the flow structure and the existence of preferential at low flow rate. The flow heterogeneity due to the coupling between the network geometry and the non-linear flow properties of the carbopol solution is better evidenced by looking at the PDF of the velocity displayed in Fig.~\ref{fig:fig8}. They contrast with the ones obtained with the Newtonian fluid. The Gaussian distribution previously observed becomes an exponentially declining distribution, and the standard deviation increases when the flow rate is decreased. These distributions demonstrate that in some channels the flow is close to zero and is much larger than the average velocity in others. The quantitative analysis of such distributions is undoubtedly a relevant tool to study the evolution of the localization of the flow, but a quantitative and detailed study of this phenomenon is beyond the scope of the present article. 


\section*{Discussion}

In the previous sections, we have shown the validity and the potentiality of the auto-correlation method combined with Schlieren contrast enchancement to map the velocity field in a complex microfluific device. Let us now discuss and details a few technical aspects which could benefit to future users. 

Let us first emphasize that the Schlieren contrast enhancement is a powerful costless alternative to fluorescence optical methods. Although fluorescence microscopy is now a standard technique, it has some technical drawbacks and contraints. For instance, it requires suitable illumination devices and filters. The Schlieren contrast enhancement only requires a grid and a standard camera, and can achieve particle detection even when the tracer size is smaller than that of the pixel. Of course, the presence of the grid alters the quality of the raw image, and induces some optical aberration as discussed in the experimental section. On the opposite, if the grid size becomes too small or if the spatial resolution is low in comparison with the mesh size, all the light deflected by the presence of the bead will be collected by the same pixel, thus cancelling the contrast enhancement effect. We therefore recommend to use a grid which wavelength is in the order of the pixel size. By doing so, the images are more uniform and do not suffer from the above mentioned aberration. Standard sieves are good solutions to obtain grids of various wavelength.

The auto-correlation method is very robust, as it could be used on poor quality images, contrary to PIV or PTV methods. The second advantage is related to the spatial resolution which is 1 pixel. Although the other methods could achieve this high resolution, they required averaging over a long period of time. In this paper, we use 3000 successive images to reach a precision of about 5\%. The precision is reduced when using less images, but we obtain satisfactory results with only 500 images. As for other velocimetry methods, the frame-rate needs to be adjusted to the magnitude of the velocity. Indeed, the characteristic time measured is inversely proportional to the velocity. In order for the auto-correlation function decay to be measured correctly, we recommend to adapt the frame rate at about a few times $1/\tau_0$. Higher values would lead to a better precision, but would required a larger number of images. Although we restrict this study to low velocities (around 10 $\mu$m/s), the method could be applied without additional difficulties to larger velocities 

Tracer size should also be adapted to the experiments. Although large tracers lead to better optical constrast, they also reduce the spatial resolution and induce finite size bias when their size approaches a fraction of that of the channels. In this work, we used tracer sizes that are much smaller (6 $\mu$m or 20 $\mu$m beads) than the channel width, and smaller or around the pixel size. The tracer concentration should also be adjusted since low concentrations requires long acquisition time, and high concentrations influences the decorrelation time due to multiple particles. We thus recommend to adjust the concentrations so that the mean distance between tracers is approx. the channel height. Finally, the refractive index of the particles has a strong influence on the contrast and on the amplitude of the intensity fluctuations. Using water with a small amount of carbopol ($n_1=1.33$\cite{Harvey1998}) instead of glycerol ($n_1=1.47$\cite{Hoyt1934}), we increased the difference between the refraction index of the tracer ($n_2=1.49$ for PMMA \cite{Michela1985}) and of the fluid. We then observed an increase in the light intensity contrast that made it considerably easier to analyze the signals of the experiment displayed on Fig.\ref{fig:fig6} as opposed to the data displayed on Fig.\ref{fig:fig4}.

The main drawback of the autocorrelation technique is that it requires a calibration procedure. The calibration factor $a$ is a length scale which depends on the experimental parameters. Since the measured characteristic time corresponds to the passage of a tracer in front a pixel. it should be - in principle - for tracers larger than the pixel size close to the tracer diameter or to the pixel size otherwise. For the data presented in this paper, we used 6$\mu$m beads and pixel size of about 20$\mu$m. The characteristic size $a$ found from velocity calibration is 10~$\mu$m which is around both the bead and pixel sizes, which is thus coherent with the previous argument. We have tested several other experimental conditions. When the pixel size is much smaller than the beads, the value of $a$ is clearly close to the bead diameter: we obtain $a=25.3 \mu$m, with 20 $\mu$m beads and a pixel size of 0.17 $\mu$m. Increasing the pixel size leads to a more complex picture. With 6 $\mu$m beads, we have increased the pixel size, starting from 20$\mu$m. We find that $a$ increases from 10 to 18~$\mu$m when the pixel size is 40$\mu$m, but then saturates when the pixel size is increased further. Possibly the reason for saturation is related to the fact that for large pixel sizes, multiple beads could be observed at the same time on the same pixel. As the precise value of $a$ depends on experimental conditions in addition to the details of characteristic time definition (here, $\tau_0$ has been defined as $C(\tau_0)=0.5)$, we think the calibration step could not be avoided to achieve quantitative measurements. However, the value of the bead size could serve as an a priori estimation for the characteristic time. 

\section*{Conclusion}

In this work, we have shown that large scale anemometry with high resolution could be achieved efficiently and using time auto-correlation of the light intensity. The method is robust since it does not require high quality imaging of the tracer particles. In this work, we have used a Schlieren technique to reinforce the contrast of the image and to achieve velocity mapping in bright field with beads smaller than the pixel size. The combination of the two methods is thus of great interest to map velocity fields in large complex microfluidic network. We have applied it to flows in micromodels of porous media, where high resolution velocimetry is required to achieve quantitative description of the phenomena. In addition to the experimental validation of the method using Newtonian fluids, we report preliminary results obtained with yield stress fluids which unambiguously shows that the flow concentrates at low flow rates in preferential paths. The technique reported here thus opens the road towards quantitative studies of complex fluid flows in model porous media. 

Beyond the applications just mentioned, we are also convinced that this technique may be useful to study the flow structure of active fluids \cite{Gachelin2014,Lopez2015,Creppy2015}. For instance, bacteria are characterized by refractive index ($n_2 \approx 1.39$ \cite{Aleksei2003,Liu2014}) slightly different from the suspending fluid (often a water solution with $n_1 = 1.33$, and have a typical size of a few microns. Our technique may then be useful to study the spatial organization of the velocity field of these fluids without the need to add tracer particles.
 

\section*{Acknowledgements}

The authors would like to thank J.-P. Hulin, D. Salin, L. Talon and T. Chevalier for useful discussions and the Agence Nationale de
la Recherche for financial support of the project LaboCothep No. ANR-12-MONU-0011.

\section*{Appendix}

The pore network modelling used to compute the theoretical PDF of the velocity inside the heterogeneous channel network consists in assuming developpped flow profiles in all the rectangular channels. We use a similar network as for the experiments, i.e. rectangular channels of uniform $h$ but of heterogeneous width $w_{ij}$. These are set randomly according to a log-normal distribution having the same standard deviation as for the experiments. 
The channels are connected through nodes of the network where mass conservation written for node labelled $i$ simply reads 
\begin{equation}
	 \sum_{j} Q_{ij} =0,
\end{equation}
where $Q_{ij}$ is the flow rate in the channel connecting node $i$ and $j$, and where the summation is made on the neighboring nodes. The flow rate $Q_{ij}$ is given for a Newtonian fluid by\cite{MORTENSEN2005}
\begin{equation}
	Q_{ij} = \frac{P_i-P_j}{\eta l}\frac{8 h^4}{\pi^3 } \sum_{n=1,3,5,...}^\infty \frac{w_{ij}}{h \pi n^4}- \frac{2}{\pi^2 n^5}\tanh\left(\frac{n \pi w_{ij}}{2 h}\right),
\end{equation}
where $\eta$ is the fluid viscosity, $l$ the length of the channel and $P_i$ the pressure at node $i$. This forms a complete set of linear equations of unknown $P_i$ that we solve numerically using matlab. 

Knowing the pressure field, we then compute the mean velocities in all the channels to obtain the PDF displayed in Fig.~\ref{fig:fig6b}.

\bibliographystyle{rsc}

\end{document}